\documentclass[journal=jacsat,manuscript=article]{achemso}

\usepackage{chemformula} 
\usepackage[T1]{fontenc} 
\usepackage{hyperref}

\author{Ayana Ghosh}
\email{ghosha@ornl.gov}
\affiliation[1]
{Center for Nanophase Materials Sciences, Oak Ridge National Laboratory, Oak Ridge, TN 37831, USA}
\alsoaffiliation[2]
{Computational Sciences and Engineering Division, Oak Ridge National Laboratory, Oak Ridge, TN  37831, USA}
\author{Gayathri Palanichamy}
\affiliation[SRMISTNANO]
{Department of Physics and Nanotechnology, SRM Institute of Science and Technology, Kattankulathur - 603 203, Tamil Nadu, India}
\author{Dennis P. Trujillo}
\affiliation[1]
{X-Ray Science Division, Argonne National Laboratory, Lemont, IL 60439, USA}
\author{Monirul Shaikh}
\affiliation[SRMISTNANO]
{Department of Physics and Nanotechnology, SRM Institute of Science and Technology, Kattankulathur - 603 203, Tamil Nadu, India}

\author{Saurabh Ghosh}
\email{saurabhghosh2802@gmail.com}
\affiliation[SRMISTNANO]
{Department of Physics and Nanotechnology, SRM Institute of Science and Technology, Kattankulathur - 603 203, Tamil Nadu, India}

\title[An \textsf{achemso} demo]
 {Insights into cation ordering of double perovskite oxides from machine learning and causal relations }

\abbreviations{IR,NMR,UV}

\begin{document}

%
%
%
%
%
\begin{abstract}
This work investigates the origins of cation ordering of double perovskites using first-principles theory computations combined with machine learning (ML) and causal relations.
We have considered various oxidation states of A, A$^{\prime}$, B, and B$^{\prime}$ from the family of transition metal ions
to construct a diverse compositional space.
A conventional framework employing traditional ML classification algorithms such as Random Forest (RF) coupled with 
appropriate features including geometry-driven and key structural modes 
leads to highly accurate prediction ($\sim$98\%) of A-site cation ordering.
We have evaluated the accuracy of ML models by entailing analyses of decision paths, assignments of probabilistic confidence bound, and finally
introducing a direct non-Gaussian acyclic structural equation model to investigate causality.
Our study suggests that the structural modes are the most important features for classifying layered, columnar and rock-salt ordering.
For clear layered ordering, the charge difference between the A and A$^{\prime}$ is the most important feature which in turn depends on the B, B$^{\prime}$ charge separation.
Based on the outputs from ML models, we have designed functional forms with these features to derive energy differences forming clear layered ordering.
The trilinear coupling between tilt, rotation, and A-site antiferroelectric displacement in Landau free-energy expansion becomes the necessary condition behind the formation of A-site cation ordering.
\end{abstract}

\section{Introduction}
Cation ordered double perovskites of the form AA$^{\prime}$BB$^{\prime}$O$_6$ with A as an alkaline-earth or rare-earth ion, B and B$^{\prime}$, as transition metal ions, exhibit a wide range of properties due to their structural and compositional flexibility\cite{solana2016double, solana2021}.
In particular, a substantial number of compounds within this family of materials as reported by both theoretical and experimental studies, 
show multiferroic and polar metallic behavior\cite{Serrate_2006, PhysRevB-CrOs, CoOs-JACS-2013, FeOs-Chem, CoOs-PRB-2015}. 
In AA$^{\prime}$BB$^{\prime}$O$_6$, the BB$^{\prime}$ sublattices typically order in rock-salt while AA$^{\prime}$ sublattice can order 
in layered [L], rock-salt [R], or columnar [C] ordering leading to diverse structural, electronic and magnetic properties\cite{VASALA20151,NiOs-PBR-2016, PhysRevM-CMSO}.
Fixing BB$^{\prime}$ as rock-salt while considering all three possible AA$^{\prime}$ orderings such as AA$^{\prime} $layered (A[L]B[R]) ordering and 
AA$^{\prime}$ rock-salt ordering (A[R]B[R]) 
lead to non-centrosymmetric space group (\textit{P}2$_1$ and \textit{Pc}, respectively), if (a$^-$a$^-$c$^+$) distortion is imposed.
Here, (A[L]B[R]) ordering is of particular interest due to the microscopic polarization arising because of
non-cancellation of layered polarization in two successive AO and A$^{\prime}$O layers.
The low symmetry phase is established by the hybrid improper ferroelectric (HIF) mechanism\cite{ebpgnature, cjfnbprl, cjfadv, mulder2013turning, cjfreview, Ghosh2015, Shaikh2020}.
The central question still remains as  how to achieve A-site cation ordering in double perovskites.
\par 
The disorder tendency at A-sites is more pronounced than at B sites.
Consequently, B-site cations tend to order more efficiently compared to the A-sites.
To form stable A/A$^{\prime}$ ordering, B/B$^{\prime}$ rock-salt ordering coupled with second-order Jahn-Teller (SOJT) distortions at the B$^{\prime}$ site  (i.e., placement of $d^0$ cations at the B$^{\prime}$ sites)
is known to be the most crucial factor\cite{wood1}.
\begin{figure*}
\centering
\includegraphics[width=\linewidth]{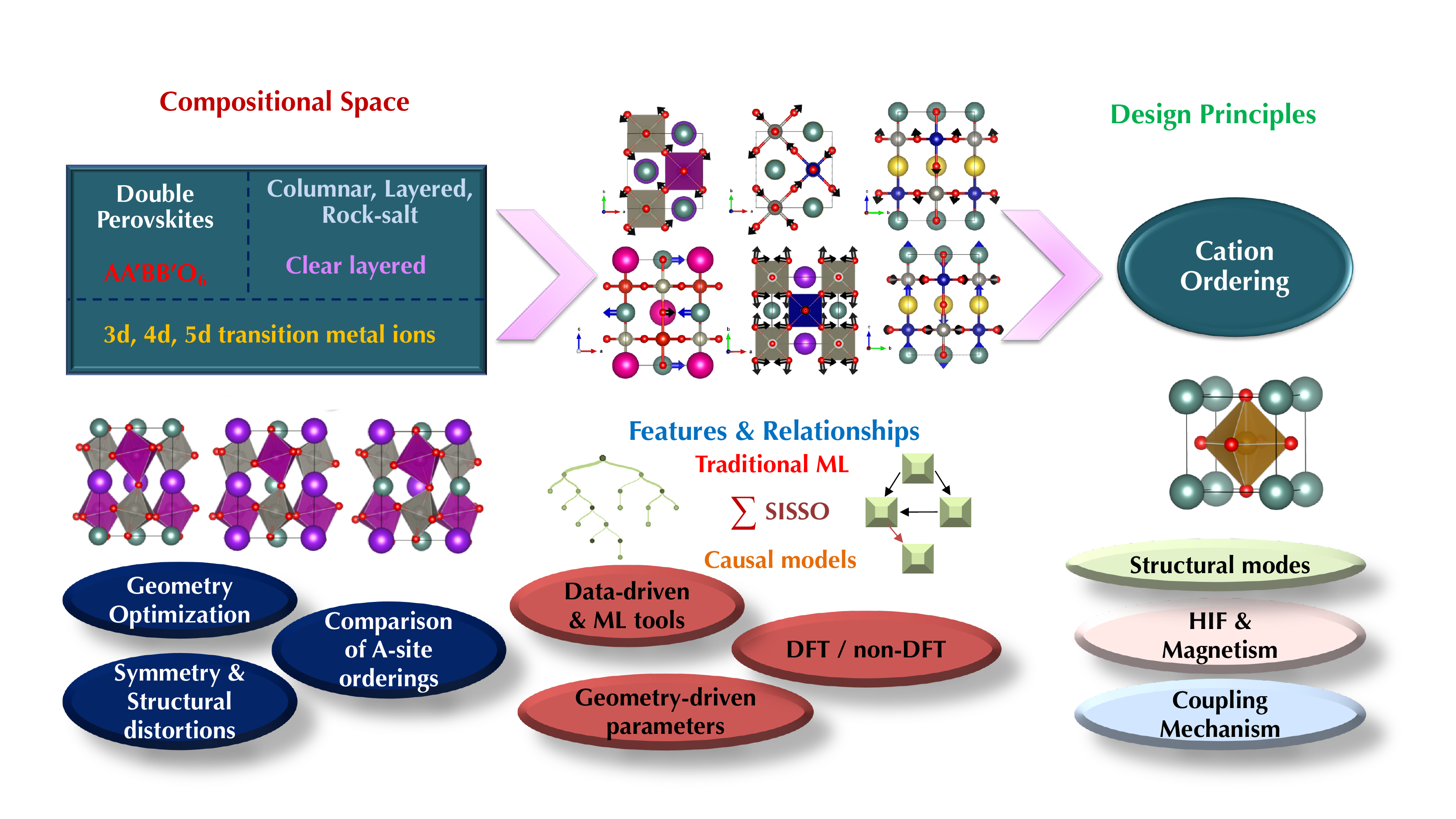}
\caption{
\textbf{Overall workflow} Flow diagram showing the primary stages of the DFT and causal ML models-based framework to gain insights on cation ordering.
The study consists of preparing a dataset using first-principles computations on selected double perovskites, evaluating structures, energies and 
multiple features.
These get utilized in data-driven techniques, ML models built using RF algorithms, respectively to explore cation ordering as well
as existing causality.
}
\label{fig:overall_framework}
\end{figure*}
In general, a clear ordering (irrespective of A-site and B-site ordering) is reported to be dependent on a variety of parameters.
A non-exhaustive list of such factors includes differences in 
cation radii and/or oxidation states, charge ordering, cooperative first order Jahn-Teller distortions of B cations (FOJT), A-site vacancies 
coupled with SOJT distortion, the tilt of BO$_6$/B$^{\prime}$O$_6$ octahedra.  
Other physical properties, structural modes, oxidation states, coordination, tolerance factor, 
external conditions and complex interplay between them can also play important roles in this context.
However, full or partial dependence on these factors can not be fully understood by sole exploitation 
of density functional theory (DFT)-based methods.
As a result, it creates a unique opportunity to evaluate if data-driven and ML techniques can help to solidify such understandings.
\par
In the community of materials science with a focus on ML, double perovskites have gained much attention in recent years.
Several reports \cite{talapatra2021machine, kim2021synthesizable, xu2018rationalizing, zhang2021machine, l2019machine, wagner2018learning, morita2022breaking, lufaso2006structure} exist on combinatorial and systematic searches conducted over a wide compositional space throughout the periodic table to judiciously shortlist formable perovskites of different forms such as 
ABO$_3$, A$_2$BB$^\prime$O$_6$, AA$^\prime$B$_2$O$_6$, and AA$^\prime$BB$^\prime$O$_6$, ABX$_3$ with energy-related applications, including solar cells, \cite{yilmaz2021critical} photovoltaics. \cite{l2019machine,howard2019machine}
The design capability of thermodynamic stability, formation energy, and synthesizability are explored using high-throughput screening, first-principles computations combined with various ML algorithms \cite{tao2021machine}.
Out of different electronic properties, bandgap and its tuneability have been the key functionalities of interest due to their direct relations with the applications mentioned above. \cite{ramprasad2017machine, chen2020critical, pilania2021machine, tao2021machine, pilania2016machine, halder2019machine, liu2017materials, butler2018machine, li2008formability, zhang2007structural, filip2018geometric, travis2016application, balachandran2018predictions}
Our study of cation ordering falls right in this regime where we are interested in formation of cation ordering driving functionalities such as switchable polarization and magnetization in AA$^{\prime}$BB$^{\prime}$O$_6$, by comparing energies between different A-site ordering.
At its current stage, these data on different A-site cation ordering are not available in the literature.
Most of the experimental observations have reported compounds where A-site layered ordering is favorable. \cite{de2018nonswitchable, pn2021structural, shankar2020site, pn2021synthesis, pn2021factors, pn2021switchable}
Thus our computational data can be a good addition for performing further explorations.
The study is intended towards adding interpretability as well as discovering cause-effect relationships (going beyond the standard practices to uncover structure-property relationships) between features representing cation ordering in double perovskites.

\par
In this work, we have employed DFT computations and supervised traditional machine learning (ML) techniques to explore such insights. 
Different possible oxidation states of A, A$^{\prime}$, B, B$^{\prime}$ with B, B$^{\prime}$ elements belonging to 3\textit{d}-3\textit{d}, 
3\textit{d}-4\textit{d} and 3\textit{d}-5\textit{d} blocks, are considered to construct datasets ranging over 
a wide compositional space. 
To design a robust set of descriptors, both DFT-derived and non-DFT-derived features are taken into consideration.
While the first contains information retrieved directly from the first-principles computations, the latter relies on 
independent (computed using non-expensive physical models) representation of structures and order parameters.
The models based on non-DFT features provide the user an alternative to obtaining a reasonable prediction of 
cation ordering in the absence of accessibility
to more robust DFT-driven features.
Comparison between these two types of models, as included in this work,
also gives a measure of the importance of the DFT-driven features for deriving reliable predictions.
Additional details on the construction of the parent dataset along with considered descriptors are listed in the Methods section.
\par
Multiple variations of the parent dataset are created to carefully examine the role of the entire descriptors space
driving the cation ordering and corresponding energy differences between different types of ordering.
A series of classification and regression models are constructed utilizing
a decision tree-type RF algorithm.\cite{scikitlearn}
While a set of models perform multiclass classification into columnar (0), layered (1), and rock-salt (2) ordering, other models
predict distinguished labels signifying the formation of clear layered ordering (0 or 1).
These models show high accuracy (balanced accuracy scores $>$94\%) for both classification and regression models.
The accuracies vary between the choice of descriptors between DFT and non-DFT features.
\par
Reasonable predictions of cation ordering along with providing signatures of 
driving factors for cation ordering can be achieved utilizing such ML models.
However, if we would like to understand what is causing the cation ordering to be of a specific kind, 
for e.g., columnar versus rock-salt and how one feature can affect the other leading to the target, assessing
linear or non-linear correlations are not enough.
In addition, from the overall accuracy score, it is hard to determine if all systems present in the dataset are being predicted
accurately with the same confidence bound.
%
\begin{figure*}
\centering
\includegraphics[width=\linewidth]{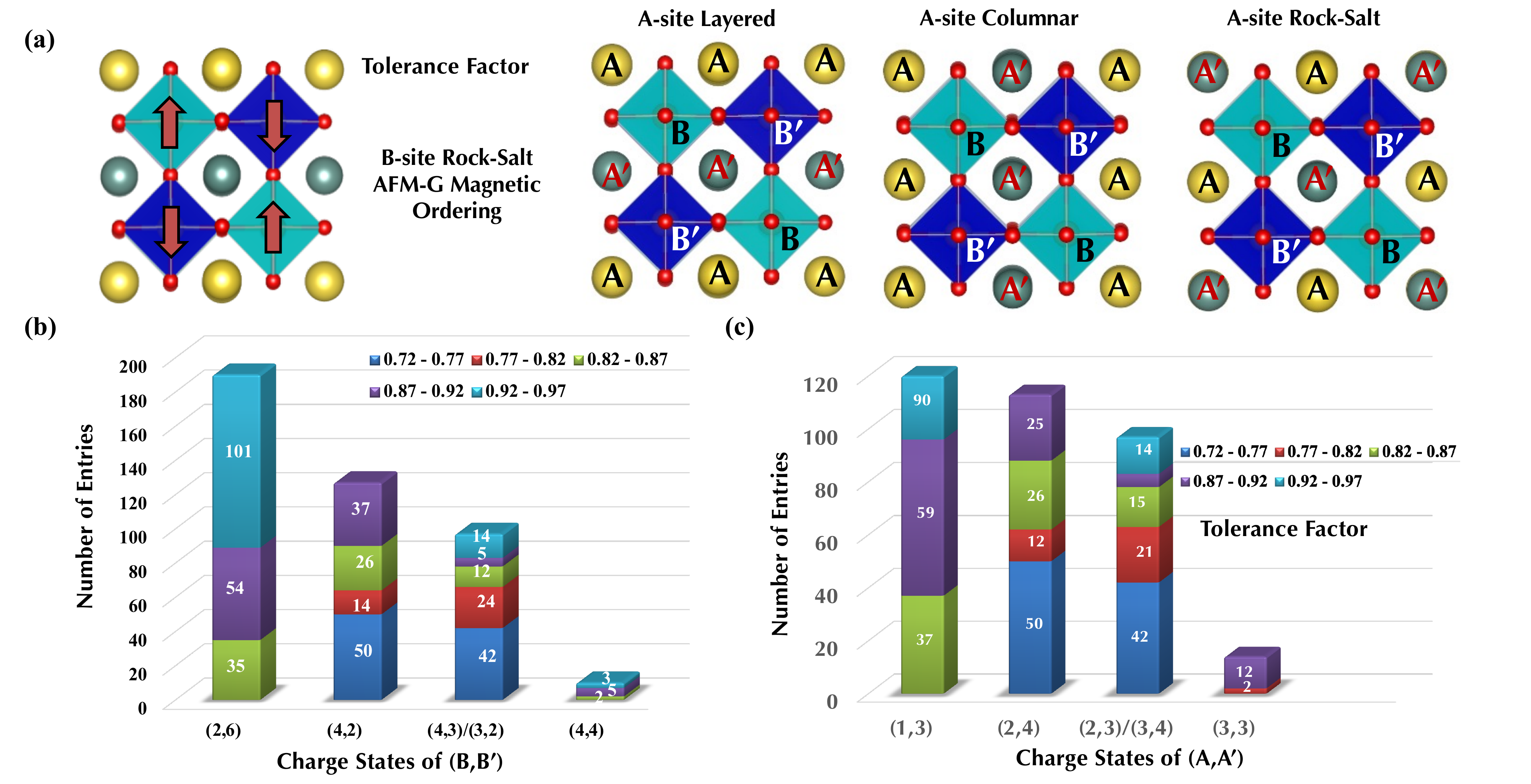}
\caption{
\textbf{Representation of double perovskites of interest}
(a) Structural illustrations of various A-site cation orderings such as
A-site layered, A-site columnar and A-site rock-Salt 
with B-site fixed as rock-salt ordering.
The magnetic configuration considered here is of G-type AFM ordering.
(b) Distribution of the data set in terms of various possible oxidation states available for BB$^{\prime}$ sites such as (2,6), (4,2), (4,3)/(3,2), and (4,4), respectively.
(c) The number of entries present in the dataset if the charge states of AA$^{\prime}$ vary as (1,3), (2,4), (2,3)/(2,4), and (3,3), respectively. 
The variations of the charge states for the systems are plotted here with respect to ranges of tolerance factor.
}
\label{fig:dist_compounds}
\end{figure*}
Hence, we have established methodologies to systematically address these potential caveats present in conventional ML
frameworks.
We have assigned probability estimates as confidence bounds for each system by considering decisions made by each decision tree
in RF.
These also indicate the existence of possible competing phases of structures which are explained in detail as a part of the study.
Next, from the combination of identified important descriptors, 
SISSO (sure independence screening and sparsifying operator) method~\cite{ouyang2018sisso, ghiringhelli2015big, ghiringhelli2017learning} 
is implemented to explore if functional forms of these primary descriptors can be formulated into linear equations to quantify the feature dependencies.
We have also visualized the decision paths (representative case can be found in Supplementary Material) to open up the black box of an ensemble of trees to show the importance of competing features regulating
the predictions of the models.
Finally, direct linear causal networks are built to study existing structure-property relationships going beyond the standard practices of correlation.
Such models delineate how multiple features are causally related, and can be tuned in different possible combinations, as a result of which, 
the target property, such as cation ordering, can be observed or modified.
\par
The key stages of the entire framework involve dataset compilation using DFT computations, evaluation of descriptors space, 
construction of interpretable ML models, assessment of functionalized features, and causality 
has been illustrated in Figure \ref{fig:overall_framework}.
All details about the associated methodologies
can be found in the Methods section.
Overall, this DFT-based study combined with causal ML models provides a comprehensive understanding of the determining 
factors behind cation order ordering exhibited by double perovskites.
\begingroup
\begin{table}
\centering 
\caption{List of \textit{d} block elements, corresponding charge states, their combinations, and examples of individual compounds for which DFT computations are performed.}
\label{table:indv_comp}
\begin{tabular}{ll}
\hline
BB$^\prime$ & \\
\hline
Elements & 3\textit{d}-3\textit{d}: Mn-Ni  \\
& 3\textit{d}-4\textit{d}: (V, Cr, Fe, Co)-(Zr)  \\
& 3\textit{d}-5\textit{d}: (V, Cr, Mn, Fe, Co, Ni)-(W, Re, Os) \\
Charge states &  3\textit{d}-3\textit{d}: Mn(+3)-Ni(+3), Mn(+4)-Ni(+2)\\
& 3\textit{d}-4\textit{d}: +4, +4\\
& 3\textit{d}-5\textit{d}: +2, +6 \\
Total charge state of (BB$^\prime$) & 3\textit{d}-3\textit{d}:: +6 \\
& 3\textit{d}-4\textit{d}: +8 \\
& 3\textit{d}-5\textit{d}: +8\\
\hline
AA$^\prime$ & \\
\hline
Elements & Group IA: Na, K, Rb \\
& Group IIA: Mg, Ca, Sr, Ba  \\
& Lanthanides: La, Ce, Nd, Sm, Gd, Tb, Dy, Ho, Tm, Lu  \\
& Others: Hf, Pb, Zn, Hg, Bi, Y, Sn, Zr, Cd \\
Charge states & Group IA: +1  \\
& Group IIA: +2 \\
& Lanthanides: +3 \\
& Zn, Hg, Sn, Cd: +2 \\
& Bi, Y: +3 \\
& Hf, Pb, Zn: +4 \\
\hline
Combination of charge states & 3\textit{d}-3\textit{d}: BB$^\prime$ = +6, AA$^\prime$ = +6 \\
& 3\textit{d}-4\textit{d}: BB$^\prime$ = +8, AA$^\prime$ = +4 \\
& 3\textit{d}-5\textit{d}: BB$^\prime$ = +8, AA$^\prime$ = +4 \\
\hline
Number of compounds & 3\textit{d}-3\textit{d}: 81\\
& 3\textit{d}-4\textit{d}: 4\\
& 3\textit{d}-5\textit{d}: 60\\
\hline
Examples & 3\textit{d}-3\textit{d}: BaDyMnNiO$_6$, CaHfMnNiO$_6$, CdZrMnNiO$_6$ \\
& 3\textit{d}-4\textit{d}: RbTmCoZrO$_6$,  NaYCrZrO$_6$ \\
& 3\textit{d}-5\textit{d}: NaYMnReO$_6$, KGdFeOsO$_6$ \\
\hline
\end{tabular}
\end{table}
\endgroup
%

%
\par
\section{Methods}
This section is divided into three primary parts such as (a) Compositional space and computations, 
(b) Features Space and (c) ML methods.
These subsections describe the compounds of interest, and methodologies adapted to perform DFT computations followed by how we have constructed datasets on a wide variety of double perovskite compounds by combining them with a various sets of features.
Brief descriptions of all ML methodologies as utilized in the study are also included in the last subsection.
\subsection{Compositional space and computations}
\textbf{Compositional space:}
In this study, we have considered various pairs of transition metal ions placed at 
A, A$^{\prime}$, B and B$^{\prime}$ cation sites.
In a formula unit cell of AA$^{\prime}$BB$^{\prime}$O$_6$ double perovskites, the oxidation states of A, A$^{\prime}$, B and B$^{\prime}$ should sum up to +12.
To ensure we investigate a wide variety of compounds, we have selected elements from 3\textit{d}, 4\textit{d}, 5\textit{d} blocks.
 We then combine them based on their corresponding charge states, coordination numbers, and tolerance factors.
A list including the specific elements from \textit{d} blocks along with their corresponding charge states, combinations of those to form AA$^{\prime}$BB$^{\prime}$O$_6$ are included in Table \ref{table:indv_comp}.
A total of 145 individual compounds are computationally designed for which density functional theory (DFT) computations are performed 
for all three types of A-site cation ordering.
The structures of three A-site cation orderings with B-site kept fixed as rock-salt ordering with G-type antiferromagnetic (AFM) configuration are illustrated 
in Figure \ref{fig:dist_compounds}(a).
The cations located at B and B$^\prime$ sites belong to 3\textit{d}-3\textit{d}, 3\textit{d}-4\textit{d} and 3\textit{d}-5\textit{d} blocks. 
Out of the total 145 compounds, 81 systems belong to 3\textit{d}-3\textit{d} BB$^\prime$ with (AA$^\prime$, BB$^\prime$) charge states as (+6, +6), 4 compounds belong to 3\textit{d}-4\textit{d} BB$^\prime$ with (AA$^\prime$, BB$^\prime$) charge states as (+8, +4) and 60 compounds belong to 
3\textit{d}-5\textit{d} BB$^\prime$ with (AA$^\prime$, BB$^\prime$) charge states as (+8, +4).
The choice of particular magnetic configuration at the B, B$^\prime$ and its dependence are additionally detailed in the later part (\underline{subsection}: DFT methodology \& assumptions) of this section.
The distributions of all charge states, tolerance factors, and evaluations of their corresponding difference in charges are further discussed in the following \underline{subsection} of Datasets construction. \\
%
%
\begin{figure*}
\centering
\includegraphics[width=1.0\textwidth]{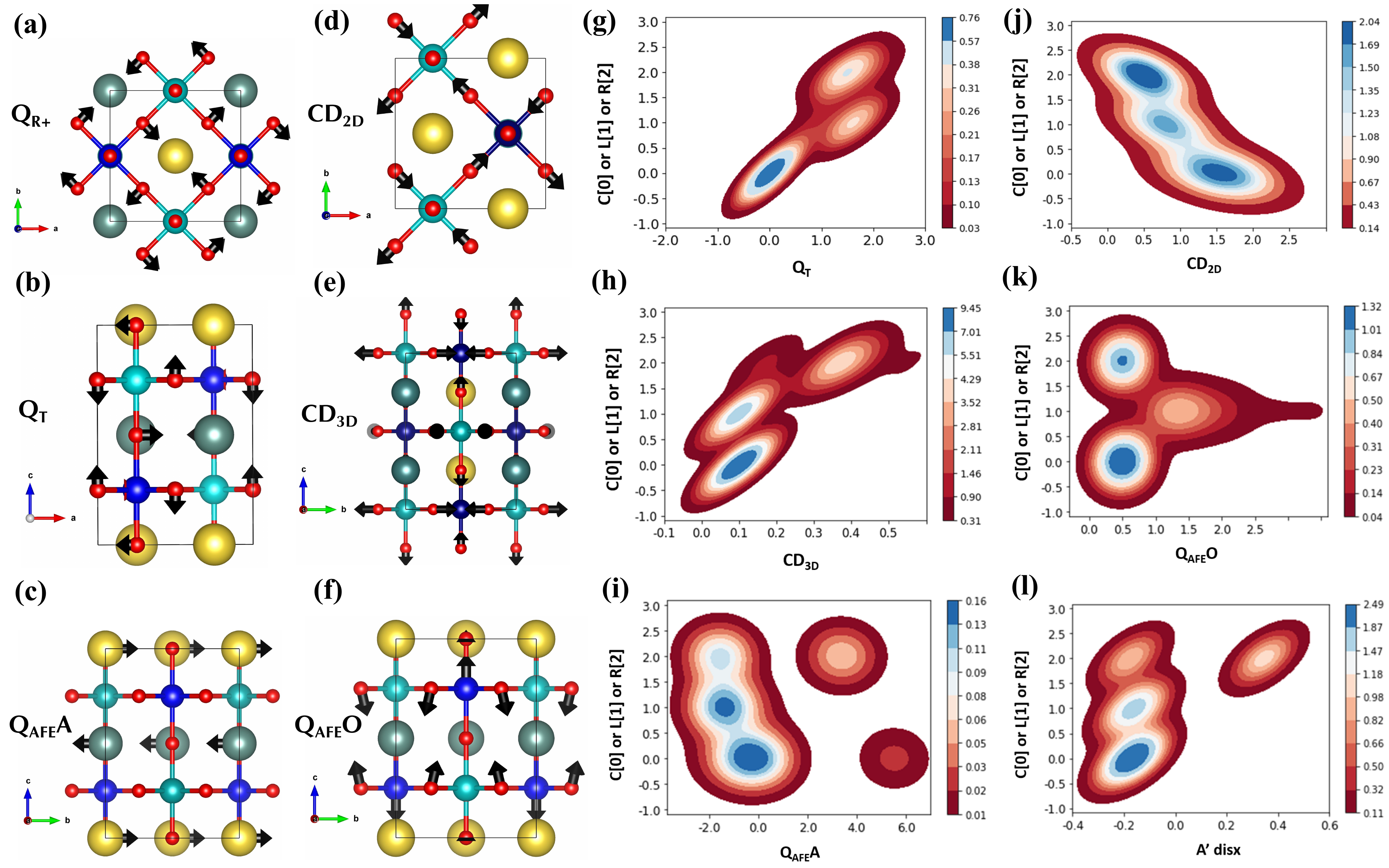}
\caption{
\textbf{Representation of structural modes and corresponding distributions}
Key structural modes such as in-phase rotation (Q$_{R+}$), tilt (Q$_T$),
antiferroelectric A-site displacement (Q$_{\text{AFE}} A$), 
2-dimensional charge disproportionation (CD$_{2D}$), 3-dimensional charge disproportionation (CD$_{3D}$)
and antiferroelectric O-site displacement (Q$_{\text{AFE}} O$), as computed for the systems are illustrated in (a-f), respectively.
In case of out-of-phase rotation (not shown here), the in-plane oxygen atoms located at both top and bottom layers of 
BO$_6$ (or B$^{\prime}$O$_6$) octahedra rotate in the opposite directions.
The 2D kernel-density plots (g-k) are utilized here to show different amplitudes of structural modes with respect to three types of A-site cation ordering.
The distribution of A$^\prime_{\text{dis}}x$ is also plotted (l) here.
}
\label{fig:str_modes}
\end{figure*}
\textbf{DFT methodology \& assumptions:}
First-principles calculations are carried out using density functional theory (DFT)\cite{dft1} with projector augmented wave (PAW) 
potentials\cite{paw} and within generalized gradient approximation (GGA) with U$_{\text{E}}$\cite{ldau1}, using Vienna ab initio simulation package (VASP)\cite{vasp}. 
The exchange-correlation part is approximated by PBEsol functional~\cite{pbesol, pbesoloverpbe}. 
Brillouin zone integrations are performed with a $\Gamma$-centered 6$\times$6$\times$4 $K$-point mesh following the crystal symmetry. 
A cutoff energy of 520 eV is set for all of the calculations, and spin polarization is taken into account. 
All relaxations are carried out until changes in the total energy between relaxation steps are within 1$\times$ 10$^{-4}$ eV and atomic forces on each of the atoms are smaller than 0.01 eV/ \AA. 
G-type AFM ordering (collinear spins) at BB$^\prime$ sites are assumed for all computations.
We have also compared the ground state energies for different magnetic configurations such as AFM-A, AFM-G, AFM-C and FM for each A site layered, rock-salt and columnar orderings (keeping B site fixed as rock-salt) for selected compounds.
These compounds are curated based on their wide variety of tolerance factors, and charge states such that the parent dataset can be well-represented.
The results (details in Supplementary Material) suggest that for 3\textit{d}-5\textit{d} combinations in AA$^{\prime}$BB$^{\prime}$O$_6$, the lowest energy magnetic ordering is AFM-G irrespective of the possible A-site cation ordering.
It is hard to conclude what is the correct magnetic ground state, just by comparing the energy differences for AA$^\prime$BB$^\prime$-(3\textit{d}-3\textit{d})O$_6$ systems due to possible disorder tendencies at the A-site.
These findings are in reasonable agreement with previous reports.\cite{MSChemMat,MResearchExp2020}
We have further investigated the effects of disorder tendencies for A/A’ sites on ground-state energies for selected compounds. 
Additional details are included in the Supplementary Material on these energies for all magnetic configurations as mentioned above as well as differences in the ground state energies between the ordered and disordered phases for these selected compounds.
The disordered phases are constructed utilizing special quasi-random structures (SQS) as suggested by A. Zunger et al.\cite{zunger1990special}
Since the energy difference between various magnetic orderings at B/B$^\prime$ is not fundamentally related to the classification of the A-site cation ordering, our assumptions and predictions using AFM-G type magnetic configurations remain valid for this study.
We have used an effective ($U$$_{\text{E}}$ = $U$$-$J$_{\text{H}}$) Hubbard parameter to account for on-site Coulomb interaction for 3\textit{d} states of transition metals such as V($U_{\text{E}}$= 2.0 eV), Cr($U_{\text{E}}$ = 3.0 eV), Mn($U_{\text{E}}$ = 4.0 eV), Fe($U_{\text{E}}$ = 4.5 eV), Co($U_{\text{E}}$ = 5.0 eV), Ni ($U_{\text{E}}$ = 6.0 eV) and 5\textit{d} states of transition metals W($U_{\text{E}}$ = 0.0 eV, Re($U_{\text{E}}$ = 1.0 eV), Os($U_{\text{E}}$ = 1.0 eV), respectively. 
The choice of $U_{\text{E}}$ is determined by the representation of correct oxidation states and convergence of local magnetic moments at the BB$^\prime$ sites in the double perovskites. \cite{kumar2010theoretical,SGPhysRevLett,MSChemMat}
Additional details are provided for each combination of BB$^\prime$, corresponding $U$$_{\text{E}}$ values, local magnetic moments, oxidation states along with energy differences between different A-site cation ordering as a function of $U$$_{\text{E}}$ can be found in the Supplementary Material.
The AMPLIMODES (symmetry mode analysis)\cite{Kirov:ks0200} and PSEUDO\cite{Kroumova:ks0096} are used to determine the magnitude of the structural modes.
\\
\textbf{Datasets construction:}
The parent dataset is constructed using the DFT computations (145$\times$3) performed on 145 compounds with three A-site cation ordering, giving rise to a total of 
435 entries.
For each compound, there are three respective entries corresponding to the three types of orders, C, L, and R.
The class labels such as 0, 1 and 2 are assigned based on the initial and final structural configurations associated with the same.
For a compound entry with label 2 for example, the final structure of the compound must resemble the template of A-site rock-salt order as shown in Figure ~\ref{fig:dist_compounds} (a).
The structural decomposition could not be achieved due to structural complexities for some of these entries.
Hence, out of all transition metal ions after mixing elements from different blocks,
the total number of compounds that are considered from the combinations of 3\textit{d}-3\textit{d}, 3\textit{d}-4\textit{d} and 3\textit{d}-5\textit{d} are 224, 10 ,and 176, respectively.
These form the parent dataset consisting of a total of 410 data points.
The distribution of all 410 entries in the dataset is plotted with respect to variations in the tolerance ranges.
 In Figure \ref{fig:dist_compounds}(b) this distribution is shown for different oxidation states for BB$^{\prime}$ sites, as (2,6), (3,2), (4,2), (4,3) and (4,4), respectively. 
For AA$^{\prime}$, the possible oxidation states become (1,3), (3,4), (2,2), (3,2) and (4,4) as represented in Figure \ref{fig:dist_compounds}(c).
The charge differences (CD) between B and B$^{\prime}$ thus evaluated are 4, 1, 2, and 0, whereas, CD between 
A and A$^{\prime}$ are 2, 1, and 0, respectively.
As an example, for a CD of 2 between A, A$^{\prime}$, where A is Na(+1) and A$^{\prime}$ is Y(+3), a total of 176 systems are considered assuming the CD between B, B$^{\prime}$ is 4 (within 3\textit{d}-5\textit{d} elements).
A subset of this dataset is also chosen to examine the formation of clear layered ordering.
For this case, a cut-off energy difference of -32 meV is set.
This choice is driven by the fact that double perovskite systems most likely exhibit clear layered ordering 
as validated by experimental observations \cite{de2018nonswitchable, shankar2020site, pn2021structural, pn2021synthesis, pn2021factors, pn2021switchable}.
More compounds may show clear layered ordered phases if the cut-off energy is lowered and vice-versa.
We specify the cut-off at -32 meV for the following reasons.
Energy such as -32 meV corresponds to room temperature.
It means if a compound satisfies this cut-off energy criterion, the formation of mixed phases for these compounds is eliminated at room temperature.
The ordering temperatures of the perovskite compounds considered in this study are not known.
Maintaining this criterion also ensures that even if the ordering occurs at a higher temperature, the energy difference between various ordered phases 
will still be that of 300 K.
As a result, the formation of any mixed phases at the time of determination of clear layered ordering can be discarded at any given temperature.
The reference energy to evaluate the energy difference between various orderings and assign corresponding labels, 
is always considered to be that of the layered ordering.
For example in the latter subset of data entries, if the energy difference between layered and columnar or rock-salt for a compound is above -32 meV, the compound is assigned to label 1.
In addition, this specific cut-off energy difference also complies with previous DFT-based investigation~\cite{MSChemMat}
that reports the formation of cation-ordered polar phases based on such criteria.
The distribution of the compositional space as shown in Figure \ref{fig:dist_compounds} represents a 
wide range of charge distributions of the cation sites.
In addition, this compositional span also encompasses a broad variety of structures ranging from cubic to orthorhombic as quantified by the differences in tolerance factors. 
Consequently, this leads to notable changes in the features space (geometry-driven features and structural modes) which are 
considered key descriptors in the construction of ML models.

\subsection{Features space}
\textbf{Geometry-driven features:}
The geometry-driven features such as charge states (C$_A$, C$_{A^{\prime}}$, C$_B$, C$_{B^{\prime}}$), 
coordination numbers (Cn$_A$, Cn$_{A^\prime}$, Cn$_B$, Cn$_{B^{\prime}}$), 
ionic radii (r$_A$, r$_{A^{\prime}}$, r$_B$, r$_{B^{\prime}}$),
average ionic radii (AA$^{\prime}_{\text{avg}}$, BB$^{\prime}_{\text{avg}}$),
tolerance factors (TF), optimized energies (Energy), Fermi energies, differences in energy (E$_{\text{diff}}$)
cell parameters along a, b and c directions (cell length$_a$, cell length$_b$, cell length$_c$), 
cell angles ($\alpha$, $\beta$, $\gamma$), cell volume,
total magnetic moments (total mag), individual magnetic moment (mag$_B$, mag$_B^\prime$, mag$_O$), 
\textit{s}, \textit{p}, \textit{d} occupancies at all cation sites (A$_\textit{s}$, A$_\textit{p}$, A$_\textit{d}$, A$^\prime$$_\textit{s}$, 
A$^\prime$$_\textit{p}$, A$^\prime$$_\textit{d}$, B$_\textit{s}$, B$_\textit{p}$, B$_\textit{d}$, B$^\prime$$_\textit{s}$, 
B$^\prime$$_\textit{p}$, B$^\prime$$_\textit{d}$,
 antiferroelectric displacements of A, A$^\prime$ sites (A$_{\text{dis}}x$, A$^\prime$$_{\text{dis}}x$, A$_{\text{dis}}y$, 
 A$^\prime$$_{\text{dis}}y$) along x, y directions,
tilt ($\theta_{\text{tilt}}$) and rotation ($\theta_{\text{rot}}$) angles are first included in the descriptors space.
All of these features are obtained utilizing the information given by the initial structures 
followed by the optimized ones within the evaluations using first-principles computations.\\
\textbf{Structural modes:}
The next segment of features space comprises of key structural modes.
Bulk double perovskites undergo phase transition from high symmetry to low symmetry driven by unstable 
phonon modes, as temperature is lowered. 
The modes driving any specific property is referred to as functional mode. 
Either single or multiple modes can be responsible for transition from high symmetry to low symmetry phase.
Hence, these modes are also proven to be well representation of existing
structural distortions, rotations, tilts between different planes or octahedra cages, Coulomb interactions,
and even charge disproportionations arising due to changes in bond lengths, volumes of B or B$^{\prime}$O$_6$ octahedra.
Below we provide a brief description of all modes that are included in this study. 
The key structural modes for the double perovskites are namely
 in-phase rotation (Q$_{R+}$), tilt (Q$_T$), out-of-phase rotation (Q$_{R-}$),
antiferroelectric A-site displacement (Q$_{\text{AFE}} A$), 
antiferroelectric O-site displacement (Q$_{\text{AFE}} O$), 
2-dimensional charge disproportionation (CD$_{2D}$) and 3-dimensional charge disproportionation (CD$_{3D}$).
The structural modes are shown in Figure \ref{fig:str_modes}(a-f).
The Q$_{R-}$ and Q$_{R+}$ modes arise due to rotations of apical in-plane oxygen atoms located at both top and bottom layers of 
BO$_6$ (or B$^{\prime}$O$_6$) octaherdra in the same and opposite directions, respectively. 
These modes exist in the $ab$-plane.
Q$_T$ in $bc$ plane comes into play when the oxygen atoms from the top and bottom layers of BO$_6$ (or B$^{\prime}$O$_6$) octahedra move in the same direction.
The displacement of the A and A$^{\prime}$-sublattices occur in opposite directions. 
The ionic radii of A and A$^{\prime}$ are different. 
Consequently, it gives rise to a ferrielectric (FiE) distortion, very often known as anti-ferroelectric A-site displacement (Q$_{\text{AFE}} A$) mode in the literature. 
This mode is described along the $b$-axis.  
For anti-ferroelectric O-site displacement mode, 
Q$_{\text{AFE}} O$ planar O-atoms are displaced towards the higher charge state due to electrostatic Coulomb's interaction.
The volume of B$^{\prime}$O$_6$ typically increases while the same for the BO$_6$ octahedra gets reduced.
This effect is quantified by the CD$_{3D}$ mode.
For CD$_{2D}$ mode, the bond lengths between BO increase in one direction and reduce in another.
The B$^{\prime}$O bond lengths overturn the effect of BO bond lengths.
The amplitudes of all of these modes are computed within the DFT regime (details included in Methods section) 
and used to construct ML models combined with the geometry-driven features.
\par
Figure \ref{fig:str_modes}(g-k) shows continuous distributions of the amplitudes of structural modes spanning over 
the full compositional space.
It is evident from {Figure \ref{fig:str_modes}}(g) that Q$_T$ is absent for compounds with C-type ordering.
This is due to the fact that in case of C-type ordering,
oxygen atoms are located at the centre of the symmetry and can not be moved towards a particular A-cation site.
Presence of CD$_{\text{3D}}$ (h) is required to exhibit R-type cation ordering.
This mode is missing in C and L-type orderings since they are confined in two dimensions and therefore, only CD$_{\text{2D}}$ mode 
becomes relevant.
The average amplitudes of Q$_{\text{AFE}} A$ (Figure \ref{fig:str_modes}(i)) modes for C, L or R-type of orderings are in the same order of magnitude such as $\sim$1.91\AA.
CD$_{\text{2D}}$ modes (Figure \ref{fig:str_modes}(j)) are only present for C or L-type of ordering as discussed. 
Q$_{\text{AFE}} O$ (Figure \ref{fig:str_modes}(k)) is suggestive of formation of L-type of ordering.
Charge separation between AO and A$^{\prime}$O layers is responsible for Q$_{\text{AFE}} O$.
This mode is missing for C and R-type orderings which can also be confirmed by analysis of the structural mode.
In this mode, the in-plane oxygen atoms move towards the smaller A-site cations (generally one with the higher oxidation state) 
to satisfy the bonding/coordination environment.
The region with high amplitudes of A$^\prime$ displacement corresponds to rock-salt whereas, for the other two types of ordering, these magnitudes are minimal.
Moreover, from similar analyses of the datasets, we can also claim that 
Q$_{R-}$ is present for all compounds exhibiting only C-type of ordering.
Thus, Q$_{R-}$ becomes extremely important among other features to determine the presence or absence of C-type ordering.
The average amplitudes of Q$_{R+}$ modes for C, L or R-type of orderings are in the same order of magnitude such as $\sim$ 1.08 \AA.
Overall, if we consider the uniqueness of the structural modes to describe L, C or R, we can emphasis on the following points:
(a) Q$_{\text{AFE}} O$, Q$_{R-}$, CD$_{\text{3D}}$ modes are only present in L, C and R ordering, respectively and 
(b) Q$_T$ and CD$_{\text{2D}}$ modes are only absent in C and R ordering, respectively.
Hence, these information on the structural modes can play an important role in classifying L, C or R-type orderings.\\

\begingroup
\begin{table*}
\centering 
\caption{List of all models with descriptors and targets as constructed in this work. 
All features related to energy such as ground-state energy, energy per unit cell are excluded from the list.$^*$}\label{table1models}
\begin{tabular}{llll}
\hline 
Model & Number of entries in dataset & Features & Target \\
\hline
I & 410 & Geometry-driven & C[0], L[1] or R[2] \\
II & 194 & Structural modes & C[0], L[1] or R[2] \\
III & 165 & Geometry-driven \& structural modes & C[0], L[1] or R[2] \\
IV & 108 & Geometry-driven & 0 or 1 \\
V & 100 & Geometry-driven \& structural modes & 0[N] or 1[O] \\
VI & 100 & Geometry-driven$^*$ \& structural modes & Energy difference \\
VII & 100 & structural modes & Energy difference \\
VIII & 165 & non-DFT derived features & C[0], L[1] or R[2]\\
IX & 100 & non-DFT derived features & 0[N] or 1[O]  \\
\hline
\end{tabular}
\end{table*}
\endgroup
\begin{figure*}
\centering
\includegraphics[width=1.0\textwidth]{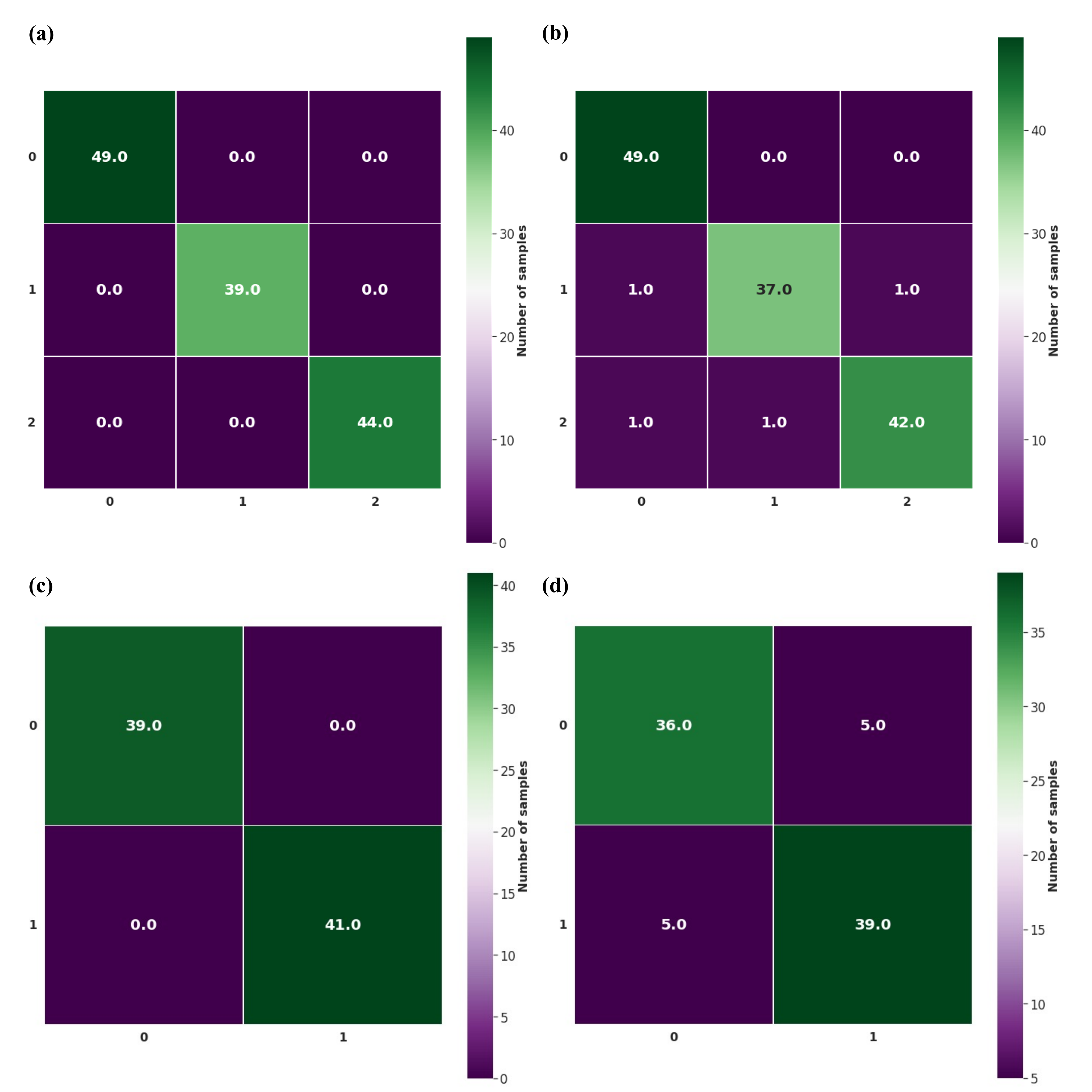}
\caption{\textbf{Confusion matrices to represent classifications of ordering for training sets}
Classification results as described by confusion matrices (a-d) for the training sets for Model III, VI, VIII and IX, respectively.
Model III, VIII use multiclass RF classification to predict columnar, layered, rock-salt ordering while for Model VI and IX, it
is a binary classification task to categorize systems in presence or absence of clear layered ordering.
}
\label{fig:confusion_matrices}
\end{figure*}
\begin{figure*}
\centering
\includegraphics[width=1.0\textwidth]{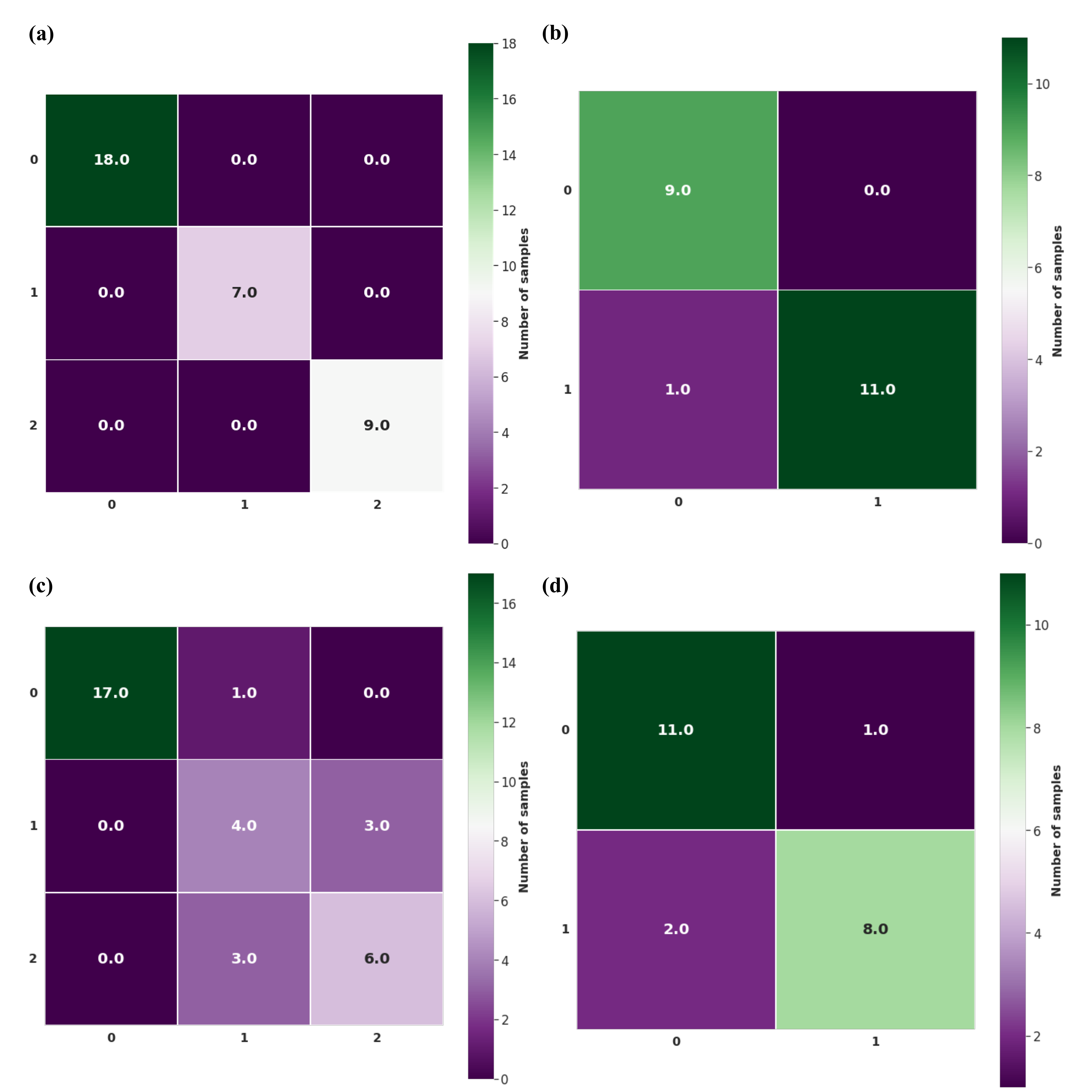}
\caption{\textbf{Confusion matrices to represent classifications of ordering for test sets}
Classification results as described by confusion matrices (a-d) for the test sets for Model III, VI, VIII and IX, respectively.
Model III, VIII use multiclass RF classification to predict columnar, layered, rock-salt ordering while for Model VI and IX, it
is a binary classification task to categorize systems in presence or absence of clear layered ordering.
}
\label{fig:confusion_matrices_test}
\end{figure*}
%

%
\begingroup
\begin{table*}
\centering 
\caption{List of the five most important descriptors for selected classification and regression models. 
}\label{table2imp}
\begin{tabular}{llll}
\hline 
Model & Type & Features (in order of descending importance) \\
\hline
II & multiclass classification & CD$_{3D}$, Q$_{AFE} A$, Q$_{R-}$, Q$_T$, Q$_{AFE} O$  \\
III & multiclass classification & CD$_{3D}$, Q$_T$, A$^\prime$$_{dis}x$, Q$_{AFE} O$, CD$_{2D}$  \\
V & binary classification & E$_{\text{diff}}$, r$_{B^{\prime}}$, C$_A$, C$_B$, A$^\prime$$_{dis}y$  \\
VI & regression & C$_B$, r$_{B^{\prime}}$, B$^\prime$$_\textit{p}$, B$^\prime$$_\textit{d}$, A$^\prime$$_{dis}y$ \\
\hline
\end{tabular}
\end{table*}
\endgroup
\textbf{Non-DFT derived features:}
Even though it is straightforward to get an estimate for most of the geometry-driven features utilizing the results from the converged computations, decomposing the phonon modes can be challenging depending on the structural complexity.
In addition, the computations may turn out to be expensive.
For elements for which reliable PAW potential may not be available, these features can not be obtained.
Thus ML models solely dependent on DFT-derived features are not enough to find a generalized model to predict cation ordering.
As an alternative, we have computed an additional set of descriptors using Automatminer \cite{dunn2020benchmarking}
to include material representations of complementary features of the DFT-derived ones.
The list of computed features is given in Supplementary Material.
This evaluation as further discussed later in the section, also leads to interesting observations stating how reliable our predictions are 
if we vary the features spaces for different compounds.
\subsection{ML Methods}
The RF classification and regression algorithms are employed to construct the ML models using Scikit-learn Python package. \cite{scikitlearn}
The choice of RF, in comparison to other classification or regression algorithms, is driven by its performance 
in the previous studies \cite{ghosh2020machine, ghosh2019assessment} when applied to datasets of small size, as well as its 
capability to rank utilized descriptors by importance.
This feature of the RF algorithms is proved to be essential for investigations where underpinning physical properties driving
the target is important.
We have considered ensemble of trees to perform the task of classifying cation ordering or predicting energy differences to form different types
of ordering.
Once the optimized models are constructed based on the best set of hyperparameters after performing full grid-search, the feature importances are
assessed to rank them based on their respective importance.
The feature importance score produced by RF represents the decrease in the weighted impurity 
over all trees for every feature.
We also note that the models results as mentioned here, were generated by averaging over 100 runs of each model. 
This is to avoid any discrepancy in the prediction results originating from utilizing small to medium dataset size for model building. \cite{LeiJCP2021}
We have introduced probabilistic confidence bounds to further evaluate the predictions from these models.
The selected features are used in the subsequent explorations to investigate the effects of functionalized descriptors using SISSO and 
existing causal ordering and strengths.
We have included additional details on the SISSO method as well as causal networks in the Supplementary Material.
\begin{figure*}
\centering
\includegraphics[width=1.0\textwidth]{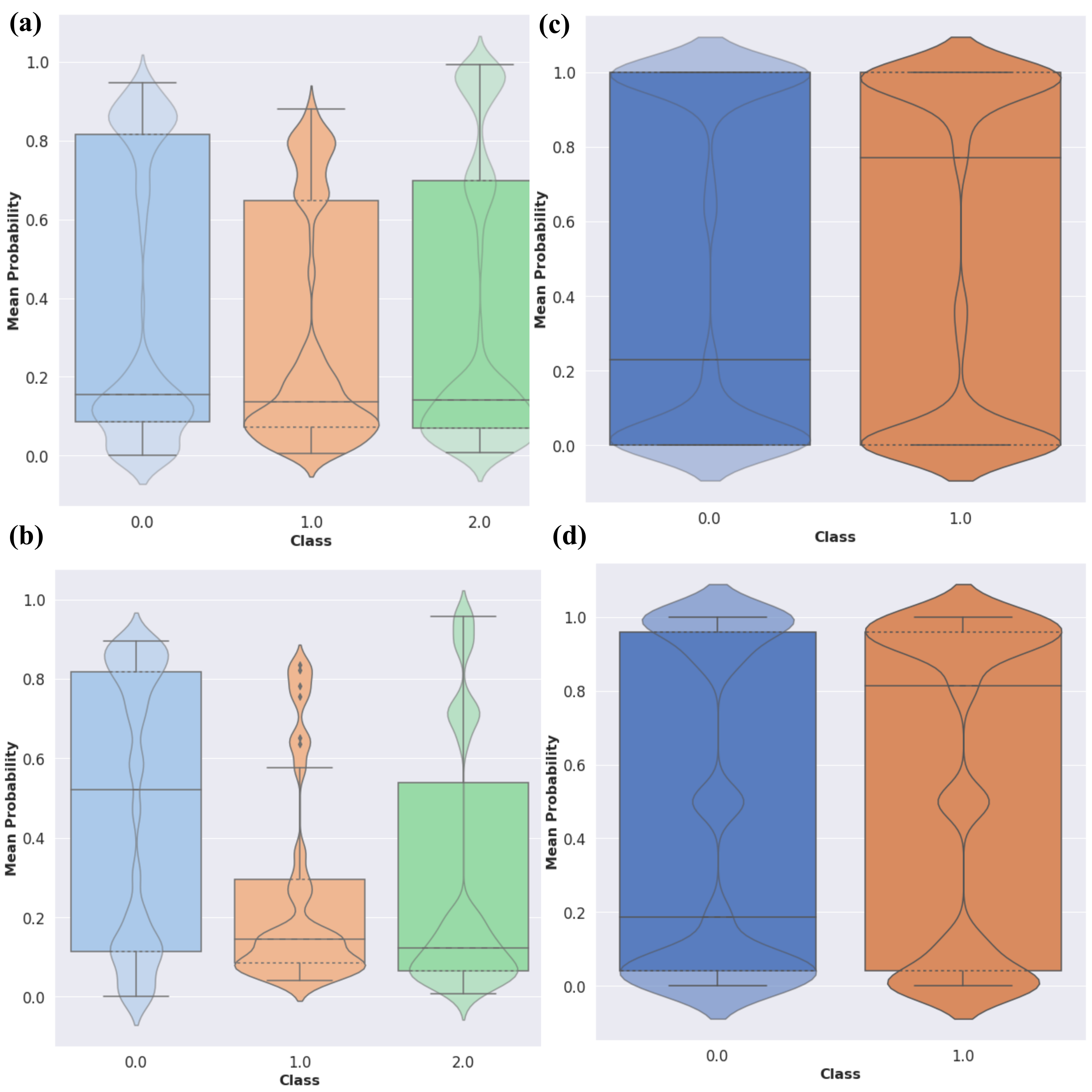}
\caption{\textbf{Mean probability estimates for predictions}
The mean probability scores (discrete as illustrated by box plots) and associated densities (continuous as illustrated by violin plots)
are shown for the training (a, c) and test (b, d) sets for Model III and VI, respectively.
These models are built with DFT features to predict columnar, layered, rock-salt, and clear layered ordering.
}
\label{fig:prob_pred1}
\end{figure*}
\begin{figure*}
\centering
\includegraphics[width=1.0\textwidth]{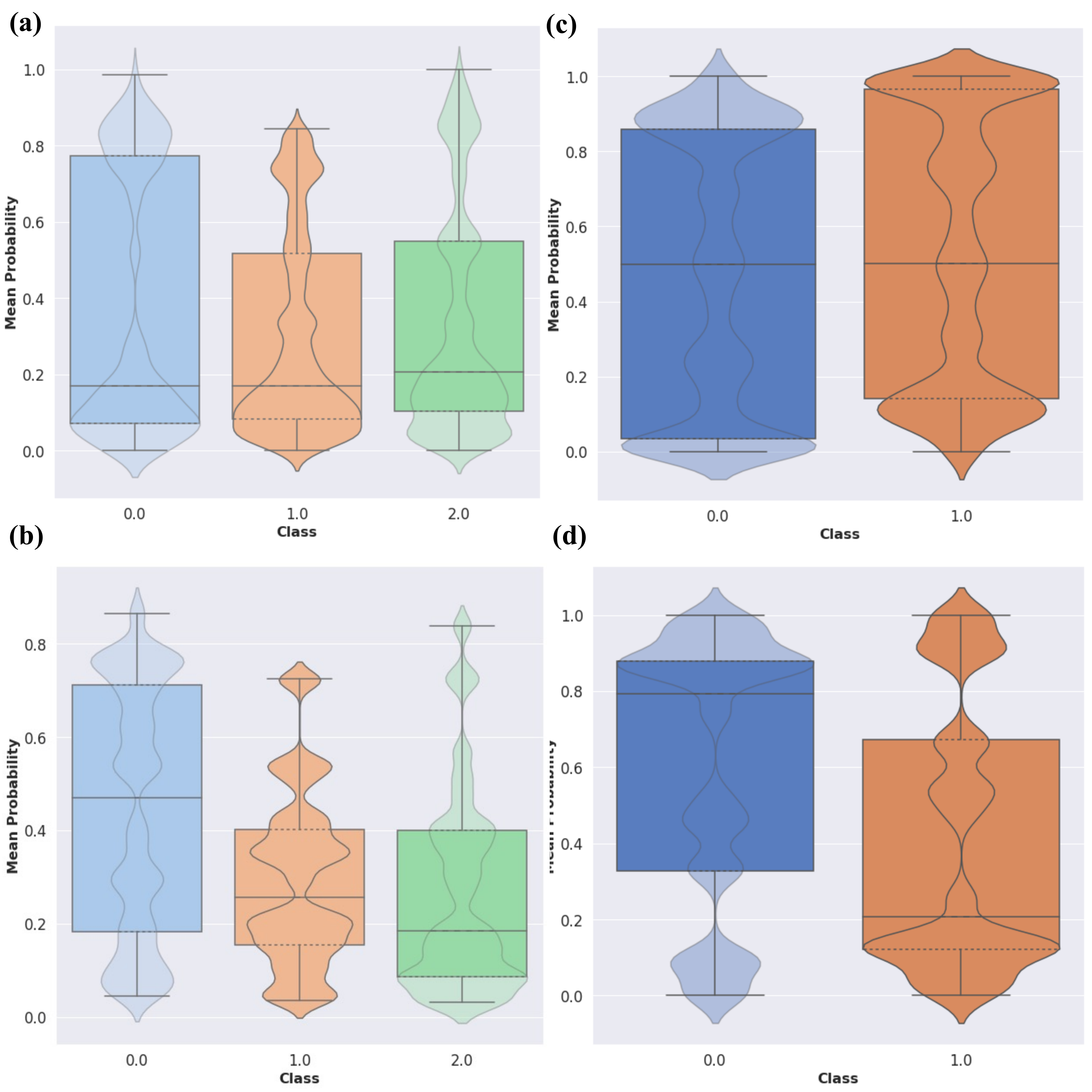}
\caption{\textbf{Mean probability estimates for predictions}
The mean probability scores (discrete as illustrated by box plots) and associated densities (continuous as illustrated by violin plots)
are shown for the training (a, c) and test (b, d) sets for Model VIII and IX, respectively.
These models are built with non-DFT-derived features to predict columnar, layered, rock-salt, and clear layered ordering.
}
\label{fig:prob_pred2}
\end{figure*}
\begin{figure}
\centering
\includegraphics[width=0.58\textwidth]{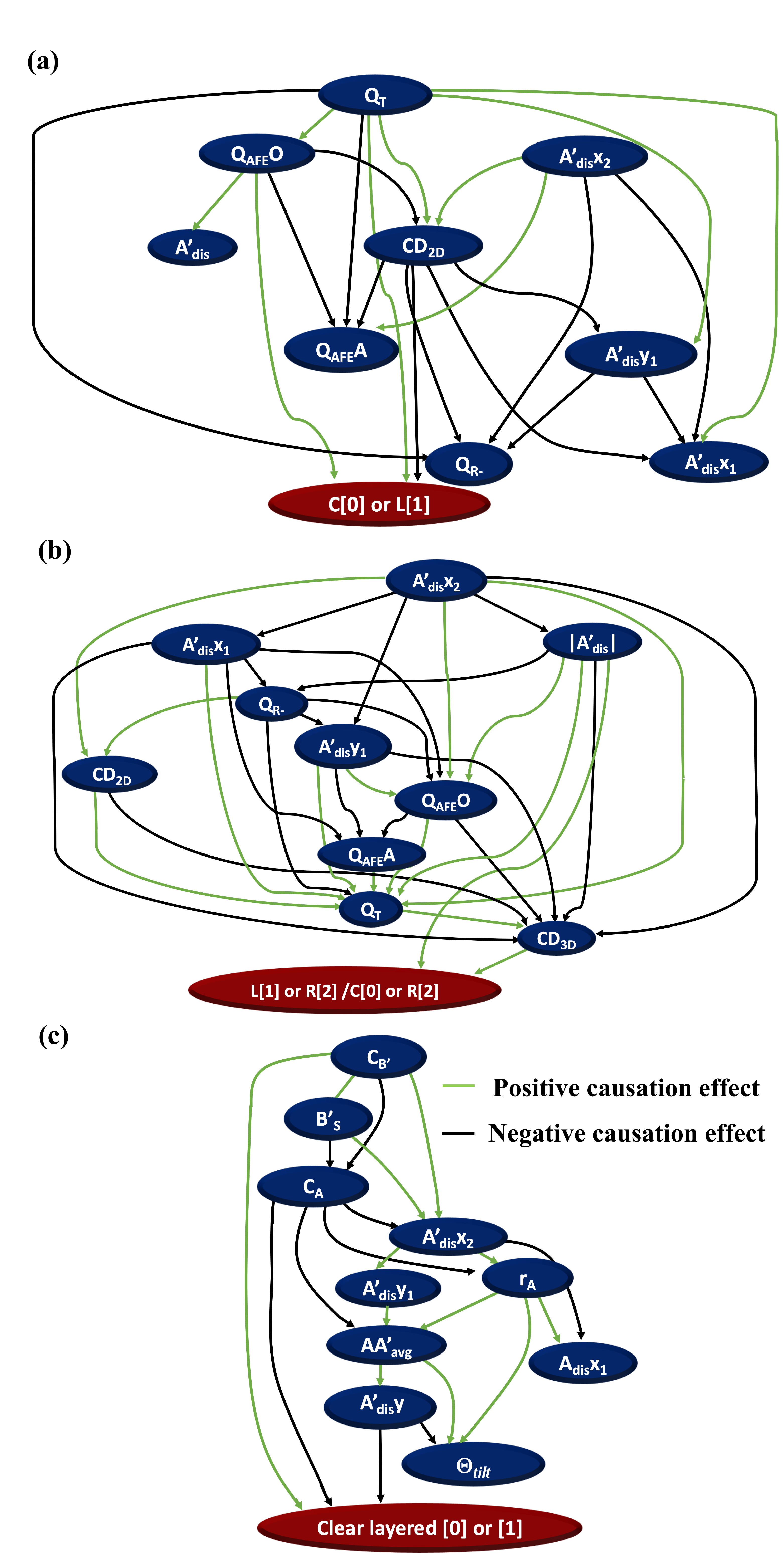}
\caption{\textbf{Direct causal networks}
Causal networks to predict three types (a,b) of cation ordering (C[0], L[1] or R[2]), and (c) clear layered ordering (0[N] or 1[O])
using ten most important features selected RF classification models constructed with descriptors space
including both geometry-driven and structural modes.
For constructing the causal networks, we have considered pair-wise classification.
In the first plot (a), the target for the causal network is to classify C and L as labeled by 0 and 1, respectively.
The causal networks to classify between L[1] and R[2] as well as C[0] and R[2] are similar as represented by (b).
Plot (c) shows the causal network to classify clear layered ordering. Here, 0[N], 1[O] denote no ordering, and ordering respectively.
}
\label{fig:causal}
\end{figure}
%

\par
\section{Result and Discussions}

\par
\textbf{RF models and physical interpretations:}
We have built a series of models by considering different combinations of the features space. 
The classification task is to either categorize compounds into (a) three types of cation ordering (C[0], L[1] or R[2]), or into (b) clear layered ordering (0[N] or 1[O]).
Energy difference to form clear layered ordering is predicted by the regression models.
A complete set of list of the models with corresponding brief descriptions can be found in Table \ref{table1models}.
Figures \ref{fig:confusion_matrices} and \ref{fig:confusion_matrices_test} represent the results of the best classification models for 
both training and test sets to predict cation ordering, respectively.
The number
of compounds for which the predicted classes match the true
classes is represented by the diagonal elements of the confusion matrix.
From the plots, it is evident that all four models as constructed by DFT-derived (a-d) and non-DFT derived features (e-h)
perform well.
The balanced accuracy scores (training, test) for these models are (1.0, 1.0), (1.0, 0.96), (0.96, 0.73) and (0.88, 0.85), respectively.
For predicting C, L or R-type of orderings, features such as CD$_{\text{3D}}$, Q$_{\text{AFE}} A$, Q$_{R-}$, Q$_T$, Q$_{\text{AFE}} O$, Q$_T$, A$^\prime$$_{\text{dis}}x$
turn out to be the most important.
Formation of clear layered ordering can be tied to descriptors such as E$_{\text{diff}}$, r$_{B^{\prime}}$, C$_A$, C$_B$, A$^\prime$$_{\text{dis}}y$.
For predicting energy differences that determine the formation of clear layered ordering, all DFT-derived features related to the energy of
the systems are discarded from the descriptors space to avoid any redundancy.
We show in Table \ref{table2imp}, that C$_B$, r$_{B^{\prime}}$, B$^\prime$$_\textit{p}$, B$^\prime$$_\textit{d}$, A$^\prime$$_{dis}y$ features 
drive the decision if a system can exhibit clear layered ordering.
This regression model shows R$^2$ of 0.93 signifying reasonable accuracy.
\par
The physical significance behind these important features can be further established by comparing these predictions with those already reported via experiments.
It has been reported that for ABO$_3$ oxides, the presence or absence of a particular structural mode (or modes) is a determining factor for identifying a particular symmetry from Raman spectroscopy studies. 
For example, in the case of nanocrystalline LuFeO$_3$ orthoferrites, it has been observed that the wavenumber of  the Raman mode $Ag(3)$ increases with the FeO$_6$ tilt angle, indicating that the Raman mode is sensitive to the orthorhombic distortion. \cite{weber2016raman,vilarinho2019crossover}
The structural modes we found are consistent with the experimental reports. \cite{blasco2021structure,ruiz2017ambient,wang2021octahedral,ezzahi2011x,hlinka2011angular,iliev2007raman,liu2019nature}
In addition, experimental evidence has been found in the presence of Q$_{\text{AFEO}}$ and Q$_\text{T}$ modes for a layered ordered double perovskite, R(R=Pr, Nd)BaMn$_2$O$_6$. \cite{blasco2021structure}
The presence of antiferroelectric A-site displacement for A-site ordered phase has been reported for BaSrM(M=Ni, Co and Mg)WO$_6$ performing Raman spectroscopy. \cite{ezzahi2011x} 
J. Ruiz-Fuertes et al. has discussed the presence of structural modes for columnar ordered CaMnTi$_2$O$_6$ \cite{ruiz2017ambient}.
There is no evidence of Q$_T$, pattern $a^-a^-c^+$, that is found in this study for this compound. 
This is exactly what we find from our predictive learning that Q$_\text{T}$ ($a^-a^-c^+$) is absent for A-site columnar ordering.
\par
For the formation of clear layered ordering, the charge separation at A- and B-sites play an important role, as derived from ML models.
These insights can be further validated by the experimental reports on ordered double perovskites NaLnNiWO$_6$ where Ln= La, Pr, Nd, Sm, Eu, Gd, and Tb. \cite{pn2021structural}
In this study, the authors have reported the effect of the difference in cationic charges on the cation ordering. 
The rock-salt ordering of the B-site cations helps to keep the highly charged cations farther apart from each other. 
In turn, it helps the AA$^\prime$BB$^\prime$O$_6$ perovskite to be completely ordered. 
Similar observations have also been reported for NaLnMWO$_6$ (M= Mn, Co, Ni, and Ln= La-Lu) for which A-site layered with B-site rock-salt ordering have been observed and stabilized. \cite{de2018nonswitchable,shankar2020site,pn2021synthesis,pn2021factors,pn2021switchable,pn2021structural}
\par
Within the regime of non-DFT derived features, components of the sine Coulomb matrix representing the electrostatic interactions 
along with the structural complexity are crucial to classify systems into C, L or R-type of ordering.
The structural complexity \cite{dunn2020benchmarking} per atom is defined as the summation of the ratio of all inequivalent sites to total sites present in the unit cell, multiplied by the
the logarithm of the same ratio.
For prediction of clear ordering, \textit{p}, \textit{d} occupancies at the cation sites along with the total number of valence electrons
turn out to be important.
We note, that these features carry full or partial information as compared to those described by the first-principles-based ones.
Thus, these counterparts are complimentary to DFT-derived ones
for systems for which DFT results become inaccessible due to either expensive nature of the calculations or unavailability of appropriate potentials.
\par
Although construction of reasonably accurate models are possible even using these set of descriptors, 
it is clear that the utilized DFT-derived features
capture the formation of cation ordering better.
The classification of C[0], L[1] and R[2] is dependent both on the presence or absence of specific structural modes in the system 
as well as on their amplitudes. 
In addition, these functional modes as driven by symmetry also affect cation displacements, cell parameters.
The rearrangement of cations in the lattice, mismatch between cation sizes are also crucial for exhibition of a particular type of A-site cation ordering.
All such information can only be well-captured by DFT computations with great accuracy.
Hence, the DFT-derived features turn out to be more relevant here.
\par
\textbf{Models accuracy and confidence bounds:}
However, even though the classification models have high overall accuracy, not all trees in the ensemble of trees may lead to same decision.
Furthermore, not all the systems can be predicted with same accuracy even within the same model as well as 
if the features space is varied.
Thus, there is a need to assign a more rigorous confidence bound as compared to using the standard accuracy measures.
Here, we compute the mean probability scores or density where discretized probability scores should add up to 1, 
for three (or two for binary classification) different classes for the multiclass classification models.
The results are represented by overlaid box and violin plots shown in Figure \ref{fig:prob_pred1} for 
the training (a, c) and test (b, d) sets of Models III, and VI respectively.
Similarly, the results are shown by overlaid box and violin plots in Figure \ref{fig:prob_pred2} for 
the training (a, c) and test (b, d) sets of Models VIII, and IX respectively.
The box plots represent the discretized probability scores for all classes.
These scores vary between 0 and 1.
The violin plots represent continuous distributions of probability density estimates.
The width of these segments show that the confidence levels to predict different classes vary, even within the same model.
A cut-off value (0.65 is considered in this study) for the probability score can be assigned to shortlist all predictions with high accuracy.
In other words, by comparing these confidence bounds, it becomes easy to distinguish the samples with high prediction 
accuracy versus that below the cut-off.
For the latter, even though the corresponding confusion matrix may show the right classification, it does not indicate the possible
existence of competing ordering which is now well-captured by this analysis.
One of the key observations derived from the analysis shows that for AA$^{\prime}$BB$^{\prime}$O$_6$ compounds, where A, A$^{\prime}$, B, B$^{\prime}$ charge states are +1, +3, +2, and +6, respectively (where B, B$^{\prime}$ are of 3\textit{d} and 5\textit{d} blocks, respectively) the formation of clear layered ordering can be predicted with high probability ($>$0.65).
However, for AA$^{\prime}$BB$^{\prime}$O$_6$ compounds with B and B$^{\prime}$ to be +2, + 4 (both B and B$^{\prime}$ are 3\textit{d} elements), the probability scores are competing, meaning high confidence level cannot be established for the predictions. 
This can further be explained if we look at the charge separation between AA$^{\prime}$ that enhances the possibility of the formation of clear ordering in 3$d$-5$d$ systems. 
The 3$d$-5$d$ systems energetically favor B-site rock-salt A-site layered cation ordered P2$_1$ phase.  
This phase is also of particular interest in terms of an exhibition of multiferroic properties in double perovskites.
\par
Utilizing the confidence bounds scheme for RF models, we have shortlisted a number of compounds for which no reports of cation ordering is available in the literature.
One list contains compounds such as BaHfMnNiO$_6$, BaPbMnNiO$_6$, BaZrMnNiO$_6$, CaCeMnNiO$_6$, HgCeMnNiO$_6$, HgHfMnNiO$_6$, NaYFeOsO$_6$, NaYFeReO$_6$.
They form A-site rock-salt ordering.
Another list is compiled including systems with A-site clear layered ordering.
The list contains compounds such as KGdCoReO$_6$, KGdVOsO$_6$, KYCrWO$_6$, KYMnOsO$_6$, NaLuCoWO$_6$, NaYCoWO$_6$, NaYCrWO$_6$, NaYVReO$_6$, NaYVWO$_6$ and RbTmVReO$_6$.
\par
To further test the robustness of the ML models, we have constructed a few additional models utilizing the dataset as used in Model III.
Thirteen randomly selected compounds are kept aside for validation and the model is retrained using the same set of geometry-driven and structural modes features.
While the training, test accuracy along with the ranking of descriptors based on importance remain the same as before, the 
balanced accuracy score for the validation set also comes out to be 1.0, all satisfying the probabilistic confidence bound.
These are the predictions on entries that this additionally-trained model has not seen before.
 Thus, it implies robustness of the classification model.
We note that dividing into training, test, and validation for all the models is ideal.
However, for a small dataset as used in this study, this segregation also takes away a significant number of entries.
Therefore, the validation scheme is only introduced here in a separate model to establish representative accuracy of the models.
\par
\textbf{Validation sets:}
(a) We have also assessed predictions on a dataset with 
5 NaYBB$^{\prime}$O$_6$ based compounds, where B = Co, Fe, Ni, and B$^{\prime}$ = Os, Re, and W. \\
Each compound in this dataset has corresponding 9 entries giving rise to a total of 45 systems.
However, as some of the structures are highly distorted and could not be decomposed with high confidence, hence we have used 41 systems for the validation data set.
The respective cation ordered phases are  CC, CL, CR, LC, LL, LR, RC, RL, RR, where, the first letter represents the ordering of the A-site and the latter is of the B-site. 
All computations assume collinear spins and G-type AFM ordering for magnetic ions. 
Apart from the reported ones, we have also found new structural modes which are relevant for capturing ordering in these combinations.
These are 2-dimensional Jahn-Teller distortion (JT$_{2D}$), 2-dimensional electronic displacement (ED$_{2D}$) and 3-dimensional electronic displacement (ED$_{3D}$), respectively (see Supplementary Material for more details). 
The new structural modes arise due to the new cation ordered phases in the validation set.
While a low accuracy of 40-50\% can be established using Model III, this performance significantly improves if the same model can
be retrained with the information on these new structural modes.
We have included additional details on these datasets and models in the Supplementary Material.
The performance on the validation sets further establishes the importance of functional modes and geometry-driven features to determine
A-site or even B-site cation ordering for a variety of systems in the same family of double perovskites. \\
(b)
Additionally, we have constructed another validation set based on selected compounds for which experimental evidence of A-site cation ordering are available \cite{king2019expanding, pn2021structural, de2018nonswitchable, shankar2020site, pn2021synthesis, pn2021factors, pn2021switchable}.\\
We have chosen a total of 17 compounds such as 
NaLaInNbO$_6$,
NaLaInTaO$_6$,
NaYMnWO$_6$,
NaYNiWO$_6$,
NaDyNiWO$_6$,
NaHoNiWO$_6$,
NaLaNiWO$_6$,
NaPrNiWO$_6$,
NaNdNiWO$_6$,
NaSmNiWO$_6$,
NaEuNiWO$_6$,
NaGdNiWO$_6$,
NaTbNiWO$_6$,
NaErCoWO$_6$,
NaTmCoWO$_6$,
NaYbCoWO$_6$, and 
NaLuCoWO$_6$.
All of these compounds crystallize into non-centrosymmetric space groups such as \textit{P}2$_1$ and \textit{P}42$_1$2 for which (a$^-$a$^-$c$^+$) distortion is possible. 
Exhibition of clear layered ordering if reported for all such compounds. 
Therefore, we have utilized Model IV which accounts for the geometry-driven features to classify compounds into clear layered category. 
All compounds in this validation set are classified correctly utilizing this model. 
\par
\textbf{SISSO:}
SISSO method allows to introduce non-linearity into the features space by searching over functionalized descriptors.
The details of this implementation can be found in the Methods section and Supplementary Material.
Overall using LASSO regression within the l$_1$ norm, optimized linear combinations of functionalized descriptors leading to the target property can be obtained. 
Table ~\ref{tableSISSO} gives the list of combinations of the functionalized features for datasets used in Model III and VI to classify into three types of ordering and predict energy differences to form clear layered ordering, respectively.
\begingroup
\begin{table*}
\centering 
\caption{List of equations for classifying A-site ordering and predicting energy differences.}
\label{tableSISSO}
\begin{tabular}{llll}
\hline 
Model & Type & Equations \\
\hline
III & multiclass classification & $\vert A^\prime$$_\text{dis}$$\vert$$^3$ + exp(-$Q_T$) + exp(-$Q_{R-}$)  \\
V & binary classification & 1/log($C_B$) + 1/log($B^\prime$$_\textit{d}$) + $1/B^\prime$$_\textit{p}$ \\
VI & regression & $C_\text{A}$$^3$ + 1/log($r_\text{A}$) + 1/log($C_{B^\prime}$) \\
VII & regression & $Q_\text{AFE}$\text{A} + $Q_{R+}$ + $Q_T^{-2}$ \\
\hline
\end{tabular}
\end{table*}
\endgroup
We only choose these four models and related important features to perform SISSO since both of these models carry the most information about the DFT-derived features.
While combinations of the descriptors may lead to the target property for both classification and regression, the accuracy of these models are comparable to those achieved using the individual descriptors.
These functional forms of the descriptors also comply with the physical intuitions regarding the structural modes and displacements of cations governing the ordering at A-site.
In particular, results derived from Model III suggests that presence of structural modes such as $Q_\text{T}$ is necessary to predict formation of columnar or layered ordering.
For rock-salt ordering, $Q_{\text{R-}}$ is necessary.
These two structural modes combined with the displacement of cations at $A^\prime$ sites are driving factors behind A-site cation ordering.
This observation is also consistent with that derived from RF models.
For binary classification and regression to distinguish between clear layered ordering as performed in Model V and VI, the differences in charge states or occupancy play a key role which is selected by the SISSO method as an optimized combination to represent clear layered ordering.
Trilinear coupling \cite{Shaikh2020} can only be established in presence of all three of the structural modes such as $Q_\text{AFE}$\text{A}, $Q_{R+}$, $Q_T$, which is a condition for the formation of clear layered ordering leading to the exhibition of tunable polarization and magnetization.
It is captured by the combination of the features derived from Model VIII predicting E$_\text{diff}$.
We note that when descriptors and functionalized forms of those are considered from a pool of geometry-driven and structural modes such as in Model VI, 
the decision is ruled by the former set.
\par
Along with establishing correlations between the features and the target by analyzing the results from SISSO, it is also possible to formulate pertinent equations to establish mathematical relationships between the features to predict physical quantity.
Below we list a couple of equations as derived to predict E$_\text{diff}$ using combination of geometry-driven and structural modes features.
\begin{equation}
\begin{split}
    \Delta E &  = \alpha (C_A^3 + \beta \frac{1}{r_A} + \gamma \frac{1}{C_{B'}})
\end{split}
\end{equation}
where, $C_A$ is the charge state of the A-site, $r_A$ is the ionic radius of the A-site and $C_{B^\prime}$ is the charge state at the $C_{B^\prime}$ site, $\alpha$ is the ground state energy (eV/5 atoms unit cell) of the system with clear layered ordering, $\beta$ is the average displacement of A and A$^\prime$ from the centrosymmetric position.
$\gamma$ is dimensionless quantity (treated as a variable) which can be optimized to predict E$_\text{diff}$.
All of these coefficients are formulated to match the overall dimensions in the energy expression.
In perovskite oxides, $Q_{AFE}A$ is proportional to $(1-\tau_{DP})$, where (DP) stands for double perovskites \cite{mulder2013turning, PhysRevB-SG-2015}.
Hence, $Q_{AFE}A = k (1-\tau_{DP})$, where k is a constant.
$\tau_{DP}$ is the tolerance factor.
The energy difference can then be reformulated in terms of the tolerance factor as proposed below.
\begin{equation}
\begin{split}
    \Delta E & =  \mu (1-\tau_{DP}) (\epsilon Q_{Tri} + \sigma \frac{Q_{R^+}}{Q_{AFE}A} + \eta \frac{Q_{R^+}}{Q_T})\\
\end{split}
\end{equation}
$\mu$ is the ground state energy (eV/5 atoms unit cell) of the system with clear layered ordering, $\epsilon$ = $\frac{1}{volume}$.
Both $\sigma$ and $\eta$ can be optimized for different material systems to obtain E$_\text{diff}$.
Plots along with details on the derivation of the equations mentioned above, showing the comparable variations in E$_\text{diff}$ as yielded by these equations and computed by DFT, are noted in the Supplementary Material.
From this type of extended analyses, we can draw further insights on formation of clear layered ordering.
The results from ML models coupled with equations derived from SISSO method suggests that the second order Jahn-Teller (SOJT) distortions due to \textit{d}$^0$ cations at the B$^\prime$ can cause additional energy gain. 
However, this is not the necessary condition for formation of clear layered ordering. 
Our study proposes that the trilinear coupling in the Landau free energy expansion is the necessary condition for clear layered ordering. 
We have also found few reports on compounds such as NaLaInTaO$_6$ and NaLaInNbO$_6$ for which A-site clear layered ordering is reported in absence of \textit{d}$^0$ cations \cite{king2019expanding}.
It further establishes confidence in our theoretical understanding as proposed in this study.
We have further investigated such connections between the features with respect to each other and the target in the later section describing causal
structure-property relationships.

\par
\textbf{Causal networks:}
Exploring feature importances using RF models already shows correlative relationships existing between specific features and the property of interest.
However, such observations are not enough to understand how the features may be driven by each other, such that it leads to the exhibition of particular ordering.
We have constructed direct non-Gaussian acyclic structural equation models (LiNGAM)\cite{shimizu2011directlingam,hyvarinen2013pairwise,liu2021exploring} to study existing cause-effect relationships among the features with respect to
different classes of cation ordering.
These models estimate causal ordering of the variables by successively subtracting the effect of each feature from the given data in the model.
The convergence is assured based on a fixed number of steps equal to the number of the variables in the model.
The ordering strengths (positive or negative) are then evaluated using conventional covariance-based methods such as least squares.
\par
For the multiclass classification into C, L or R, we visualize the causal networks (Figure \ref{fig:causal}) pairwise by considering exhibition of 
either (a) C or L, (b) L or R/ C or R and finally into (c) formation of clear layered ordering, respectively.
The positive and negative causal ordering strengths are quantified by the green and black arrows in the plots.
Interestingly, the causal paths found to predict orderings such as L or R and C or R are the same whereas for distinction between exhibition of C or L, different causal relationships are followed.
In Figure \ref{fig:causal}(a), we have shown the causation effect for C or L-type ordering. 
To predict C or L-type of ordering, Q$_T$ becomes the top feature in the causal network, sequentially driving the other descriptors such as displacements at A$^\prime$ sites and charge disproportionation modes towards the target. 
During our data distribution analysis as mentioned earlier, Q$_T$ and Q$_{\text{AFE}}$O modes are present in L ordering but absent in C. 
Thus the presence (or absence) of Q$_T$ and Q$_{\text{AFE}}$O should have a direct effect in classifying L or C. 
We are able to revisit the same findings within the cause-effect analysis . 
Q$_T$ has a positive causation effect on Q$_{\text{AFE}}$O and a negative causation effect on Q$_{\text{AFE}}$A. 
Such observations are in fact true and can be explained by crystal symmetry arguments. 
The Q$_T$, Q$_{\text{AFE}}$O and Q$_{\text{AFE}}$A distortions are shown in Figure \ref{fig:str_modes}(b), (c) and (f), respectively. 
Q$_T$, Q$_{\text{AFE}}$A and Q$_{R+}$ establish trilinear coupling in Landau free energy expansion. 
Hence, in the absence of Q$_{R+}$, Q$_T$ and Q$_{\text{AFE}}$A modes will not couple with each other leading to a negative causation effect. 
It is revealed from the analysis of the structural modes in the section, that in the case of layered ordering, coupling between Q$_{\text{AFE}}$O and Q$_{T}$ are complimentary leading to a gain in energy.
For layered ordering, Q$_{\text{AFE}}$O amounts for the displacement of planar O atoms towards a higher charge state along the pseudocubic [001] and [00$\bar{1}$] directions.
Q$_T$ mode vectors are also displaced along the same direction and constructive coupling is possible. 
In the case of Q$_{R}$, the mode vectors are displaced along [110] and [1$\bar{1}$0] directions in a complementary manner and modify the mode vectors of Q$_{\text{AFE}}$O. 
Thus, even from the symmetry point of view, constructive coupling between Q$_{\text{AFE}}$O and Q$_T$ is more feasible.
Considering Figure \ref{fig:causal}(b), it is evident that tuning the AFE displacements systematically for cations at A$^\prime$ sites can affect the disproportionation modes.
These features are crucial to establish a direct causal path to predict L or R/ C or R types of ordering.
Another viable option to predict these types of ordering follows the path of displacements but also considers the effects of Q$_{\text{AFE}} A$ and Q$_T$.
Here, we can see that all the A$^\prime$ site displacements have positive causation effects towards Q$_T$. 
In tilt distortion Q$_T$, the apical O atoms in the A$^\prime$O plane displace towards the A$^\prime$ site to establish a bonding state.
Due to the same reason, A$^\prime$ atoms displace towards O site, leading towards positive causation effect.
\par
For the formation of clear layered ordering, the information about the structural modes become less important as compared to the geometry-driven features.
Here, the direct relationships between the cation radii, average cation displacements at different A sites determine if formation of clear layered ordering is possible.
If the descriptors related to ground-state energies are included to construct another network,
multiple causal links can then be established following routes heavily dependent on these features.
This is expected given the information about the energies becomes just enough to directly distinguish between cases that may form clear-layered ordering.
In other words, the energy cut-off that is utilized to label this dataset becomes extremely relevant.
It is clear from the Figure \ref{fig:causal}(c), the charge states of A/A$^{\prime}$ and B/B$^{\prime}$ are the main decisive factors for classifying clear layered ordering.  
The charge state of B$^{\prime}$, C$_B^{\prime}$ leading to \textit{d}$^{0}$ configuration at B$^{\prime}$ site are reported to gain energy for layered ordering with respect to the competing phases. 
It also indicates formation of favorable clear layered ordering with \textit{d}$^0$ cation located at the B$^\prime$ site.
In addition, within stable B-site layered ordering, large charge separation between A and A$^{\prime}$ ( i.e., (+1, +3) AA$^{\prime}$ vs. (+3, +3) AA$^{\prime}$) is also reported to be crucial for stable layered ordering.
The displacements at the A, A$^\prime$ sites are affected by B$^\prime$.
It also suggests that the relationship between charge states and displacements together play a role in formation of clear layered ordering.
Utilizing such relationships, it is also safe to say that by increasing or decreasing the mode amplitudes of Q$_T$, CD$_{2D}$, one can directly tune the exhibition of C or L type of ordering in double perovskite systems.
Similarly, using the direct relationships between A$^\prime$$_{\text{dis}}x$ and CD$_{\text{3D}}$, one
can determine if L or R type of orderings can be formed.
For clear layered ordering, the charge state of B$^{\prime}$ and A (or A$^{\prime}$) become the governing factors.

\section{Conclusions}
%
In summary, we have employed a combination of first-principles computations and ML methods to derive insights of cation ordering in double perovskites. 
Exhibition of such ordering is assessed over a compositional space by generating structures belonging to a wide variety of crystal symmetry with
transition metal cations of all possible charge states.
The findings of the supervised models constructed with RF algorithm are substantiated 
by introducing probabilistic accuracy metrics as well as causal networks between the geometry-driven, structural modes related features and cation ordering.
We note that even though G-type AFM ordering is considered in this study, similar approach 
as presented in this work can be utilized to learn about cation orderings assuming different magnetic states. 
%
%
%
\par
Our machine learning (ML) effort accomplishes two primary goals of classifying 
(a) A-site cation ordering into columnar, layered, rock-salt ordering and, (b)
categorizing compounds into clear A-site layered ordering along with predicting associated energy differences. \\
In both cases, we have investigated the geometry-driven and structural modes related to features, and underlying mechanisms for the exhibition of a specific cation ordering using causal ML methods.
While RF algorithm allows us to construct accurate models to achieve both objectives, additional probabilistic approach helps us to shortlist candidates with different types of A-site cation ordering (for which information on cation ordering is not available before) with high confidence.
Causal networks add to the understanding of the underlying structure property relationships, going beyond the standard practices of the correlative ML models.
\par
Our investigation shows that 3\textit{d}-5\textit{d} systems form clear layered A-site cation ordering. 
The AFM-G configuration has the lowest energy for this type of systems. 
For 3\textit{d}-3\textit{d} compounds, there exists some disordering tendencies for both structural and magnetic configurations. 
The energy difference between various magnetic orderings at B/B$^\prime$ is fundamentally not related to the classification of the A-site cation ordering. 
Hence, the predictions from the ML models are expected to remain unchanged even if we consider other magnetic configurations.
\par
Below we outline the main insights derived from this study. \\
(a) Structural modes are the most important features for classifying layered, columnar and rock-salt ordering. \\
(b) In case of clear layered ordering, the charge difference between the A and A$^{\prime}$ is the most important feature which in turn depends on the B, B$^{\prime}$ charge separation.
The 3\textit{d}(B)-5\textit{d}(B$^{\prime}$) systems with A, A$^{\prime}$ in (+1, +3) charge states combination show B-site rock-salt and A-site-layered cation-ordered polar P2$_1$ phases. \\
(c) Based on the outputs from RF models and SISSO method, we are able to design equations for obtaining the energy difference to form clear layered ordering as a function of tolerance factor, trilinear coupling, charge states of A, B sites and ionic radius of the A-site cation. \\
(d) Since the energy difference is directly dependent on (1 - $\tau$), cubic symmetry does not aid in formation of clear layered ordering. In other words, 
the possibility to form clear layered ordering is enhanced for compounds with orthorhombic symmetry. \\
(e) The necessary condition to form A-site cation ordering is the trilinear coupling between tilt, rotation and A-site antiferroelectric displacement in Landau free-energy expansion.
This is contrary to the known criteria behind A-site cation ordering.
The second order Jahn-Teller (SOJT) distortions due to \textit{d}$^0$ cations at the B$^\prime$ site is only optional for exhibition of A-site cation ordering.
All of these insights are crucial information towards rational design of hybrid improper ferroelectrics, 
as achieved by utilizing a combination of ML, symmetry and DFT calculations
which can not otherwise be captured.
These are also in good agreement from the experimental studies performed on compounds from the same family of double perovskite systems, further validating our observation from the causal ML models.
\par
The framework proposed in this work allows for a couple of notable advances in the first-principles and ML community.
First is its potential to serve as a template to go beyond the traditional correlative nature of ML models to 
establish causal relationships between the descriptors space and target property within the data originated from simulations.
It also attempts to focus on the interpretability, and extensibility of the models rather than solely relying on the final predictions.
\section{Data Availability}
All datasets utilized to construct ML models can be openly accessed via the open-access link.
\href{https://doi.org/10.5281/zenodo.6570994}{DFT-computed datasets for cation ordering in double perovskites}.
\section{Code Availability}
Example implementations of RF models, confidence bounds, SISSO and causal networks can be found via open-access Colab notebooks hosted by the github repository. \\
\href{https://github.com/aghosh92/Cation-Ordering-ML}{Cation-Ordering-ML github repository}
\section{Author Contributions}
A.G. and S.G. conceived the idea of the project and wrote the manuscript.
A.G. performed, supervised all ML frameworks used and developed in this project.
G.P. and M.S performed the DFT computations under the guidance of S.G and A.G. 
D.P.T. implemented the SISSO method together with A.G.
\section{Competing Interests}
The authors have no competing interests to declare.
\begin{acknowledgement}
This effort (machine learning) is based upon work supported by the U.S. Department of Energy (DOE), Office of Science, 
Office of Basic Energy Sciences Data, Artificial Intelligence and Machine Learning at DOE Scientific User Facilities (A.G.).
Part of this research was conducted at the Center for Nanophase Materials Sciences, which is a DOE Office of Science User Facility.
A.G. acknowledges Dr. Sergei V. Kalinin (ORNL) and Dr. Maxim Ziatdinov (ORNL) for introduction to causal modeling.
AG acknowledges NERSC for providing supercomputing facility  
S.G. acknowledges DST-SERB Core Research Grant File No. CRG/2018/001728 for funding. 
M.S. thanks DST-INSPIRE (IF170335), Govt. of India for his fellowship.
G.P. and S.G. thank High Performance Computing Center, SRM IST KTR for providing the computational facility.
\end{acknowledgement}

\bibliography{SaurabhGhosh}

\end{document}